\documentclass[journal]{IEEEtran}

\usepackage{amsmath}
\usepackage{amsfonts}
\usepackage{graphicx}
\usepackage{hyperref}

\usepackage{xcolor}
\usepackage{cite}

\usepackage{booktabs}
\usepackage{comment}
\usepackage{rotating}
\usepackage{url}

\usepackage{ifpdf}

\usepackage{algorithmic}
\usepackage{array}

\ifCLASSOPTIONcompsoc
\usepackage[caption=false,font=normalsize,labelfont=sf,textfont=sf]{subfig}
\else
\usepackage[caption=false,font=footnotesize]{subfig}
\fi

\usepackage{dblfloatfix}

\ifCLASSOPTIONcaptionsoff
  \usepackage[nomarkers]{endfloat}
 \let\MYoriglatexcaption\caption
 \renewcommand{\caption}[2][\relax]{\MYoriglatexcaption[#2]{#2}}
\fi

\usepackage{url}

\begin{document}

\title{Explainable Android Malware Detection and Malicious Code Localization Using Graph Attention 
} 

\author{Merve~Cigdem~Ipek,              and~Sevil~Sen
\thanks{M. C. Ipek is with Aselsan Inc. and the Department of Computer Engineering, Hacettepe University, Ankara, Turkiye, email: ipekmervecigdem@gmail.com}
\thanks{S. Sen is with the Department of Computer Engineering, Hacettepe University, Ankara, Turkiye, email: ssen@cs.hacettepe.edu.tr}
\thanks{
This work has been submitted to the IEEE for possible publication. Copyright may be transferred without notice, after which this version may no longer be accessible.
}}


\maketitle

\begin{abstract}
With the escalating threat of malware, particularly on mobile devices, the demand for effective analysis methods has never been higher. While existing security solutions, including AI-based approaches, offer promise, their lack of transparency constraints the understanding of detected threats. Manual analysis remains time-consuming and reliant on scarce expertise. To address these challenges, we propose a novel approach called XAIDroid that leverages graph neural networks (GNNs) and graph attention mechanisms for automatically locating malicious code snippets within malware. By representing code as API call graphs, XAIDroid captures semantic context and enhances resilience against obfuscation. Utilizing the Graph Attention Model (GAM) and  Graph Attention Network (GAT), we assign importance scores to API nodes, facilitating focused attention on critical information for malicious code localization. Evaluation on synthetic and real-world malware datasets demonstrates the efficacy of our approach, achieving high recall and F1-score rates for malicious code localization. The successful implementation of automatic malicious code localization enhances the scalability, interpretability, and reliability of malware analysis.

\end{abstract}

\begin{IEEEkeywords}
Android, Mobile Security, Malicious Code Localization, Explainable AI (XAI), Graph Neural Networks, Attention Mechanism.
\end{IEEEkeywords}

\ifCLASSOPTIONpeerreview
\begin{center} \bfseries EDICS Category: NET-ATTP-MAL 
\end{center}
\fi
\IEEEpeerreviewmaketitle

\section{Introduction}
\IEEEPARstart{M}{alware} is one of the most significant cybersecurity threats today. According to Trend Micro’s 2021 Security Report~\cite{trendmicro2021}, a total of 3,315,539 malware files were detected, with approximately 77.4\% classified as unknown malware. Similarly, the 2024 Mid-Year Cybersecurity Threat Report~\cite{trendmicro2024} notes that Trend Micro blocked over 75.9 billion threats in the first half of 2024. As mobile devices have become integral to daily life, they have increasingly become targets for attackers, prompting extensive research to protect these devices from malicious threats.

The security community has invested significant effort in detecting, analyzing, and documenting such threats.  In both academia and industry, static and dynamic analysis-based security solutions have been developed and continuously improved as new and more sophisticated malware emerges on a daily basis. Artificial intelligence-based solutions, with their ability to process big data, automate tasks, and produce highly accurate results, have surpassed traditional methods. However, these approaches often function as black boxes, failing to explain why software is classified as harmful or benign. Consequently, even when an application is correctly identified as malicious, considerable time and manual effort are required to localize the malicious code and understand its behavior. Moreover, this process demands highly experienced malware analysts, which is a rare commodity in today’s world.

If malicious code could be located automatically, the time required for analysis would be significantly reduced compared to the current manual approach. With the rising global demand for malware analysts and the rapid growth of unknown malware and new variants, automating malware analysis is crucial. Achieving this would allow security professionals to shift their focus from manual code analysis to improving security measures. Furthermore, it would offer new perspectives on malware analysis and explanation, enhancing processes such as malware detection \cite{MSDroid}, malware authorship attribution \cite{gray2021identifying}, malware development analysis \cite{huang2018large}, lineage tracking \cite{cozzi2020tangled}, and automated security report generation. Consequently, malware analysis studies would become more scalable and reliable while also contributing to advancements in explainable artificial intelligence (XAI).

The proposed approach called XAIDroid aims to present a novel method for automated malicious code localization. The key hypothesis is to naturally represent code as a graph that is syntactically, lexically, and, more importantly, semantically comprehensive, and to develop new techniques for malware analysis and malicious code localization by utilizing the significant advancements made in recent years in graph neural networks (GNNs). Although the mapping between graphs and their feature vectors is deterministic in traditional studies, GNNs can help automate the understanding of these complex relations \cite{wu2020comprehensive}. XAIDroid aims to investigate the use of graph-based approaches together with an attention mechanism for the problem at hand called malicious code localization (MCL). 

XAIDroid generates API call graphs of Android applications to capture semantic relations and enhance resilience to obfuscation. Subsequently, Graph Attention Model (GAM) \cite{GAM_paper} and Graph Attention Network (GAT) \cite{velickovic2017gat, brody2021gat2}, in order to locate malicious code, are employed to assign importance scores to nodes based on their content and connections with other nodes. This approach enables focused attention on relevant information, which is crucial for tasks such as identifying high-importance nodes within the graph, an essential step in locating malicious parts of the code in our study.

Our novel approach greatly accelerates code localization, benefiting malware analysts in their tasks. For our research, we used a dataset of 7,626 benignware and 8,724 malware applications. The malware dataset includes artificially generated samples from the Mystique dataset \cite{MYSTIQUE_paper} and real malware samples from the AMD \cite{AMD} and CICMalDroid 2020 \cite{maldroid_1, maldroid_2} datasets, while benignware samples were obtained from the Google Play Store \cite{googleplay}, ApkPure \cite{apkpure}, and CICMalDroid 2020 \cite{maldroid_1, maldroid_2}. Our findings confirm the effectiveness of our approach. In malware code localization at both the class and method levels, our model achieved recall rates of 97.92\% and 97.23\%, and F1 scores of 95.75\% and 92.92\%, respectively.

The contributions of our study are summarized below.

\begin{enumerate}

    \item \textbf{Explainability:} A novel approach based on Graph Neural Networks (GNNs) and attention mechanisms is proposed to locate malicious code in a comprehensive and interpretable manner. Unlike other studies that primarily focus on malware detection using AI as a black-box tool, our method emphasizes explainable AI. By employing attention techniques, our model identifies and highlights the critical parts of the API call graphs that led to the classification of applications as malware or benignware. This transparency ensures that the reasoning behind the model's decisions is clear and actionable, representing a significant advancement in cybersecurity.

    \item \textbf{Fine-grained Localization:} XAIDroid offers malicious code localization at both the class level and the fine-grained method level. Given that research on malicious code localization is limited \cite{DROIDETEC_paper}\cite{MKLDROID_paper}\cite{dexbert} and existing studies focus mainly on class-level detection \cite{MKLDROID_paper}\cite{dexbert}, our methodology introduces a significant novelty by providing detailed method-level analysis.

    \item \textbf{Real-world Malware:} The proposed approach is evaluated on both synthetically generated malware and real-world malware, with malicious code snippets manually tagged. In contrast to existing studies such as MKLDroid ~\cite{MKLDROID_paper} and DexBERT ~\cite{dexbert}, which rely exclusively on synthetically generated malware and therefore lack validation in real-world scenarios, our method is robustly applied to real-world malware samples. Our approach addresses these limitations, offering a more comprehensive and open evaluation framework. To promote reproducibility and further research, we have made our code and dataset publicly available on our GitHub repository\footnote{\label{xaidroid_github}The source code and dataset are available at: \url{https://github.com/mervecigdem/XAIDroid}}.

\end{enumerate}

The remainder of this paper is structured as follows: Section \ref{sec:related_work} explores related work, emphasizing existing methods and identifying gaps in the investigation of malicious code localization in Android. In Section \ref{sec:the_method}, we outline our methodology, which covers dataset collection, the representation of Android applications for analysis, and our novel approach to malicious code localization. Section \ref{sec:experimantal_results} presents detailed experimental results that show its effectiveness in pinpointing malicious code at both the class and method levels within applications. Furthermore, we compare our results with recent studies, namely MKLDroid \cite{MKLDROID_paper} and DexBERT \cite{dexbert}. Threats to the validity of the proposed approach are discussed in Section \ref{sec:threats_to_validity}. Section \ref{sec:conclusion} concludes the article by summarizing our key findings and discussing the importance of our research. Section \ref{sec:ethical} outlines the ethical considerations addressed during the research process.

\section{Related Work}
\label{sec:related_work}

With advancements in GNNs, which enable code representation and processing syntactically, lexically, and semantically through graphs, GNN-based studies for malware detection are increasing. The approach in \cite{ma2019combination} generates and combines three models in control flow graphs, representing API calls by usage, frequency, and time-series data. By integrating three LSTM-based models, it achieved a 98\% detection rate. Another study \cite{pektacs2020deep} applies deep learning by converting API call graphs into lower-dimensional representations, optimizing hyperparameters to reach 98\% accuracy. GDroid \cite{gao2021gdroid} constructs a heterogeneous network linking applications and APIs, employing Graph Convolutional Networks (GCN) to achieve 97\% accuracy in malware detection and 95\% in malware family classification. AMDASE \cite{Amdase} extracts API semantics using call graphs and one-hot encoding of call pairs for feature embedding, then applies traditional machine learning techniques such as Random Forest (RF), Support Vector Machine (SVM), and K-Nearest Neighbors for malware detection. Additionally, graph-based approaches detect obfuscated malware by dynamically extracting graphs from API calls \cite{li2022dmalnet} or network behaviors \cite{fu2022encrypted}.

While many machine learning-based studies have explored Android malware detection by modeling application code or behaviors as graphs or other representations, relatively few focus specifically on Malicious Code Localization (MCL).

In Droidetec \cite{DROIDETEC_paper}, an LSTM-based model treats application code as natural language, representing it as sequences of API calls. Each execution path is captured by an API-subsequence, with no explicit connections between subsequences. An attention mechanism, commonly used in natural language processing (NLP), assigns weights to each API call by evaluating the surrounding API calls. However, the weights of API calls within a method are not summed to determine its vulnerability score, as methods with many API calls may have high weights despite containing low-priority calls. Instead, the highest-weighted k API calls are selected, and the methods invoking these calls are listed as suspicious. The scores of these methods are calculated by summing the weights of the APIs they contain, with the top n methods identified as involving malicious code. As n increases, the hit rate (probability of detecting malicious code) increases, but the accuracy decreases due to the marking of harmless code as malicious. Although Droidetec utilizes real-world samples and achieves a 91\% hit rate—calculated as the proportion of malicious samples containing at least one detected malicious code segment—its code base and data remain inaccessible.

In \cite{MKLDROID_paper}, a kernel-based graph study represents five different views of applications as graphs: API dependencies, permissions dependencies, information source/sink dependencies, instruction sequences, and control flow graph (CFG) patterns. These views address various types of malware, with trade-offs between accuracy and efficiency. However, the study lacks a representation of data flows in applications, limiting its ability to detect leak attacks \cite{MKLDROID_paper}. For feature reduction, the Contextual Weisfeiler-Lehman Kernel (CWLM) \cite{narayanan2016contextual} is used, followed by Sequence Minimal Optimization-Multiple Kernel Learning (SMO-MKL) \cite{SMO-MKL} to combine these representations for input to Support Vector Machines (SVM) for classification. The scores of code blocks are used to localize malicious code. The study differentiates malicious applications from benign ones with high accuracy using various views. However, it is based on batch learning, which reduces performance over time \cite{MKLDROID_paper}. The most effective representation is obtained from API calls. Despite achieving high accuracy on an artificially created dataset using Mystique \cite{MYSTIQUE_paper}, the approach has a high false positive rate (17\%) and its malicious code localization results require improvement. Although claimed to be fine-grained, it only evaluates performance at the class level.

DexBERT \cite{dexbert} introduces an innovative approach for localizing malicious code in Android applications by adapting natural language processing (NLP) techniques. Unlike traditional BERT, which operates on word sequences to generate context, DexBERT applies this method to Smali code, treating it as a sequence of tokens similar to words in text. The approach involves extracting Smali code and instruction sequences from each class within an APK, generating sequence embeddings using a BERT-like model, and applying classification techniques such as masked language modeling, next-sentence prediction, and auto-encoding to classify classes as benign or malicious. DexBERT focuses on class-level localization, as method-level localization is more complex due to the fine-grained nature of the code. The study uses the Mystique dataset \cite{MYSTIQUE_paper}, an artificially generated collection for model training and evaluation. While DexBERT shows promising results, its performance on real-world malware is limited, with only logic bomb detection demonstrated on a small dataset. The complexity and diversity of real-world datasets may present challenges not encountered with the Mystique dataset, potentially limiting the model's practical effectiveness.

A recent study represents applications as function call graphs (FCGs) and applies graph convolution to propagate information among neighboring nodes \cite{mcl_Android_GNN}. By including all parts of the code, including third-party libraries, they generate large graphs. They assign scores to nodes using attention, selecting the highest-scoring nodes as candidates. Despite applying multiple convolution layers to learn broader subgraph information, this goal does not seem fully achieved, as the candidate malicious list is extended by including neighboring nodes. The selection of malicious nodes is based on their use of sensitive permissions in the corresponding node/function, raising questions about the validity of node-based localization.

An earlier study focused on detecting malicious code in repackaged applications was proposed in \cite{piggybacked_li_1}\cite{piggybacked_li_2}. Their hypothesis, that all injected code is malicious, is often inaccurate, as it overlooks standalone applications. Although standalone malware constitutes approximately 13\% of all malware in the wild \cite{suarez2018eight}, this figure may be underestimated, as seen in Chinese Android markets, where only 38.3\% of malware samples are repackaged \cite{wang2018beyond}, and in the AMD dataset, where 50\% of samples are standalone \cite{wei2017deep}. The HookRanker approach, which defines metrics to identify transitions from benign to malicious behavior through method calls or system/user events, achieves 83.6\% accuracy in detecting malicious code. However, it only identifies malicious packages and does not perform localization at the class or method level. In \cite{salem2018idea}, malicious code is located by analyzing API call traces generated at runtime using the Hidden Markov Model. The trace is recursively split, and the likelihood of each segment being malicious is calculated. The study demonstrates the importance of API call proximity in detecting malicious behavior, achieving an accuracy of approximately 80\% with an average proximity of 1.6 API calls. As with any dynamic analysis-based technique, this method is limited by the need to trigger malicious code.

In summary, XAIDroid introduces a novel approach utilizing attention mechanisms and Graph Neural Networks (GNNs) for fine-grained malicious code detection, distinguishing itself from prior methods that focus on node-level attention \cite{mcl_Android_GNN}, employ deterministic graph embeddings \cite{MKLDROID_paper}, or rely on API call subsequences \cite{DROIDETEC_paper}, which lack semantic richness in graph representations. Additionally, XAIDroid offers malicious code localization at both the class and method levels, capturing essential semantic information within code behavior, thereby enhancing the effectiveness of detection and localization in Android applications.

\begin{figure*}[t]
   \centering
   \includegraphics[width=0.9\textwidth]{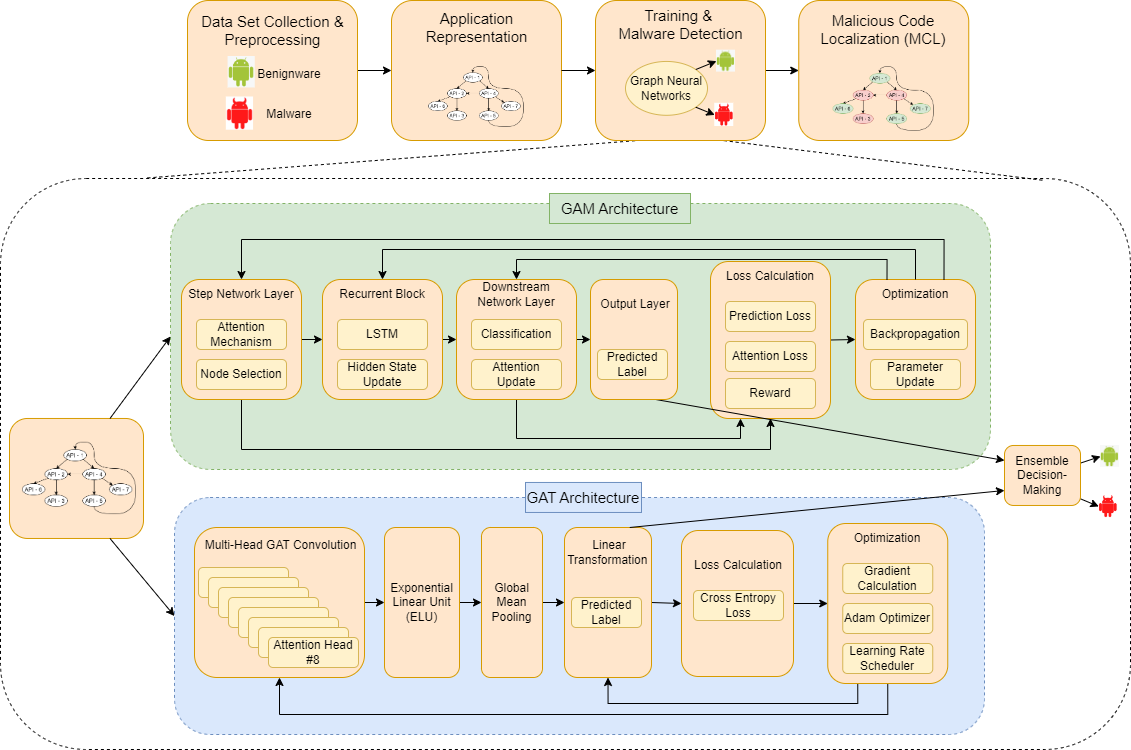}
   \caption{XAIDroid Conceptual Diagram}
   \label{fig:architecture_diagram}
\end{figure*}

\section{The Method}
\label{sec:the_method}

The objective of our research is to accurately localize malicious code within malware APK files. To achieve this, we first compile a diverse dataset from sources such as Google Play ~\cite{googleplay}, ApkPure ~\cite{apkpure}, CICMalDroid 2020 ~\cite{maldroid_1, maldroid_2}, AMD ~\cite{AMD}, and Mystique ~\cite{MYSTIQUE_paper}, encompassing both benign and malicious applications. A baseline for malicious code localization (MCL) is then established by manually analyzing malware samples. Next, these applications are converted into API call graphs, which represent the execution flow and API-level interactions. We then train Graph Attention Model (GAM) ~\cite{GAM_paper} and Graph Attention Network (GAT) ~\cite{velickovic2017gat, brody2021gat2} models using these graphs, assigning attention values to API nodes. Finally, MCL is performed at both coarse and fine-grained levels using the attention values from GAM training, with the results compared to our baseline MCL. The overall process is illustrated in Figure \ref{fig:architecture_diagram}.

\subsection{Dataset Collection and Preprocessing}

\begin{table*}[ht]
  \centering
  \caption{The Dataset}
  \label{tab:dataset}
  \begin{tabular}{cccccc}
    \toprule
    \textbf{Dataset} & \textbf{Source} & \textbf{Training} & \textbf{Testing} & \textbf{Total} & \textbf{MCL Samples} \\
    \midrule
    Benignware Dataset & GooglePlay \cite{googleplay}, ApkPure \cite{apkpure}, & 5,400 & 2,226 & 7,626 & - \\
    & CICMalDroid 2020 \cite{maldroid_1, maldroid_2} & & & & \\
    \midrule
    Trojan.AndroidOS.Mobtes.co & Mystique \cite{MYSTIQUE_paper}	& 1,850 & 797 & 2,647 & 797 \\
    Trojan-SMS.AndroidOS.FakeInst.a & CICMalDroid 2020 \cite{maldroid_1, maldroid_2} & 1,050 & 443 & 1,493 & 15 \\
    Trojan-SMS.AndroidOS.Agent.aax	& CICMalDroid 2020 \cite{maldroid_1, maldroid_2} & 750 & 318 & 1,068 & 15 \\
    Trojan-SMS.AndroidOS.Opfake.bo & CICMalDroid 2020 \cite{maldroid_1, maldroid_2}	& 700 & 304 & 1,004 & 15 \\
    Trojan-Dropper.AndroidOS.Wapnor.a & CICMalDroid 2020 \cite{maldroid_1, maldroid_2} & 690 & 289 & 979 & 15 \\
    Trojan-Spy.AndroidOS.Agent.bz & AMD \cite{AMD} & 460 & 194 & 654 & 15 \\
    Trojan-Banker.AndroidOS.Gepew.b & AMD \cite{AMD}, CICMalDroid 2020 \cite{maldroid_1, maldroid_2} & 220 & 94 & 314 & 15 \\
    AdWare.AndroidOS.Apofer.b & AMD \cite{AMD}, CICMalDroid 2020 \cite{maldroid_1, maldroid_2}  & 215 & 88 & 303 & 15 \\
    Trojan-Spy.AndroidOS.Agent.qu & Mystique \cite{MYSTIQUE_paper} & 185 & 77 & 262 & 77 \\
    \midrule
    Malware Dataset (Total) & & 6,120 & 2,604 & 8,724 \\
    \midrule
    \textbf{Total} & & \textbf{11,520} & \textbf{4,830} & \textbf{16,350} \\
    \bottomrule
  \end{tabular}
\end{table*}

To gather the benignware data set, we collected applications from Google Play ~\cite{googleplay} and ApkPure ~\cite{apkpure}, and also acquired the benignware data set from CICMalDroid 2020 ~\cite{maldroid_1, maldroid_2}. We then scanned all samples using VirusTotal ~\cite{virus_total}.  
If even one vendor flagged an application as malware, we eliminated that application from our benignware dataset. Ultimately, we retained 7,626 APK files that VirusTotal confirmed as benign. 

For the malware dataset, we utilized samples from the AMD \cite{AMD}, CICMalDroid 2020 \cite{maldroid_1, maldroid_2}, and Mystique \cite{MYSTIQUE_paper} datasets. The AMD \cite{AMD} and CICMalDroid \cite{maldroid_1, maldroid_2} datasets contain real-world Android malware, including variants such as adware, banking malware, SMS malware, and riskware. In contrast, the Mystique \cite{MYSTIQUE_paper} dataset is artificially generated, deriving 226 attack features (AF) from the MalGenome dataset ~\cite{genome} and 14 evasion features (EF) commonly used in mobile malware. Examples of such features include startup events, boot completion, and SMS reception, which trigger malicious activities. Mystique uses the Multiple-Objective Evolutionary Algorithm (MOEA) to automatically generate new malware variants, focusing on privacy leakage and modularized Android malware generation.

After acquiring these datasets, we scanned the applications using VirusTotal to determine their malware families, classes, and variants. Due to the varied notations used by different vendors in VirusTotal, we adopted Kaspersky's~\cite{kaspersky} classification system to ensure consistency. The names of Kaspersky malware ~\cite{kaspersky} generally adhere to the format \textless Type\textgreater.\textless OS\textgreater.\textless Family\textgreater.\textless Variant\textgreater. The Type identifies the nature of the threat, such as Trojan-Spy (monitoring user activities) and Trojan-SMS (sends unauthorized SMS). The OS specifies the target platform, which in this case is AndroidOS, while the Family and Variant indicate the specific strain and version within the malware family. Based on the results, we classified the malware applications according to Kaspersky's malware terminology. Our trials revealed that to effectively capture localization patterns, the training dataset for each malware variant needed to be sufficiently large. Therefore, we selected malware variants with more than 250 samples to ensure robust model performance. We identified nine such variants that met this criterion and utilized them in our analysis. 

We divided benign and malware applications into training and test sets, with approximately 70\% allocated to training and 30\% to testing. The specific malware variants and their corresponding train-test splits are detailed in Table ~\ref{tab:dataset}.

To effectively evaluate our results, with the primary objective of locating malicious code, we require malware samples with identifiable and well-documented malicious code snippets. However, such real-world samples are not available in the literature. Therefore, we manually analyzed 15 random samples from the testing set for each variant to locate malicious code. This process involved examining the Smali code of the malware and verifying malicious patterns and behaviors using static analysis results from Koodous~\cite{koodous}.

The synthetic Mystique dataset provides the necessary information, making it suitable for evaluating our approach. We utilize all samples from two malware families in the Mystique dataset to assess malicious code localization. However, Mystique only offers class-level localization. To evaluate the fine-grained capability of our approach, we manually analyze each method in the Mystique dataset to locate malicious methods. During this process, we observed that outer classes, such as "Lit/fku/juph/QYGD", were labeled as malicious, while inner classes like "Lit/fku/juph/QYGD\$1" and "Lit/fku/juph/QYGD\$2" were not. Upon closer inspection, we found real malicious behavior within these inner classes in several APKs. Consequently, we updated our classification to include these inner classes as malicious, revising the malicious class baseline from the original study~\cite{MKLDROID_paper}.

\subsection{Application Representation} 

In this study, applications are represented as API call graphs, as they provide semantic insights into the applications, display execution order, and are resilient to renaming obfuscation techniques. In these graphs, API calls are represented as nodes, with edges indicating the execution sequence. To manage graph complexity, we focus solely on sensitive API calls.

To achieve this, we utilize mappings provided by Axplorer~\cite{axplorer}, which identify sensitive API calls linked to dangerous permissions across API levels 16 to 25, offering broader coverage than Arcade~\cite{aafer2018precise}. We further expand this list by including APIs related to Java Security, Java Cryptography Extension (JCE), and Dalvik dynamic code loading libraries, resulting in a comprehensive set of 3,121 Android APIs. However, we focus specifically on sensitive APIs used in both malware and benign applications within our training set. APIs used in fewer than 10 applications are excluded, leaving a final set of 688 sensitive APIs.

\begin{figure}[t]
   \centering    \includegraphics[width=0.5\textwidth]{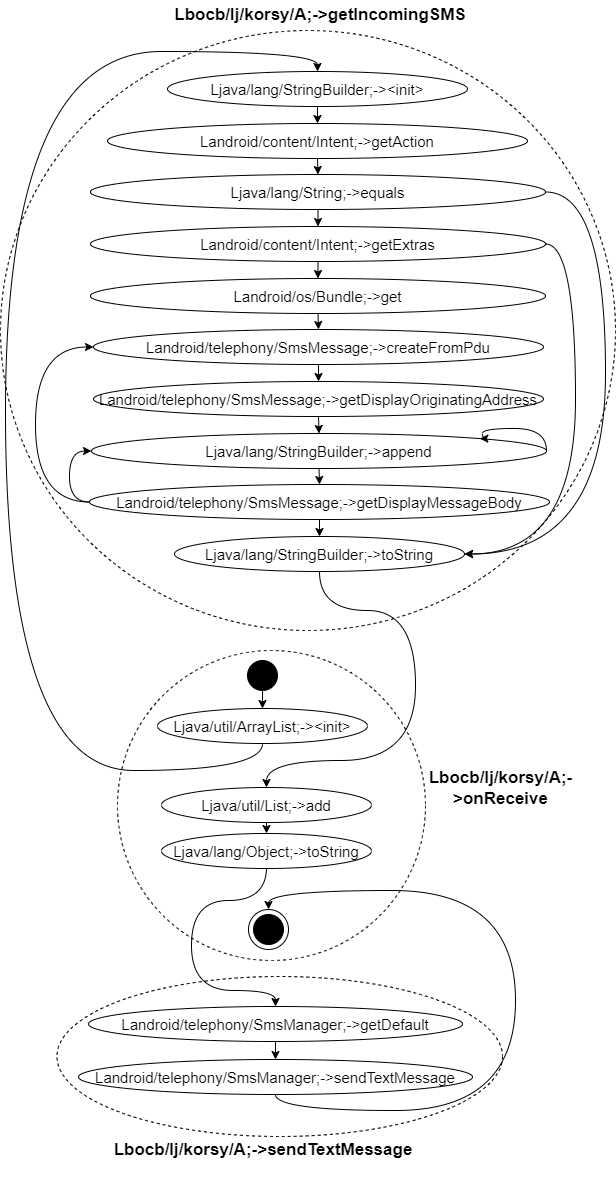}
   \caption{API Call Graph of the Smali Code given in Table \ref{tab:smali}}
   \label{fig:api_cg}
\end{figure}

In constructing API call graphs, we first build separate graphs for each method, which are then merged into a single, comprehensive graph by utilizing the calling relationships among methods. This approach avoids the need to explicitly identify entry points within Android APKs, a common challenge in program graph generation. Figure~\ref{fig:api_cg} illustrates this merging process, showing the API call graph construction for the code in Table~\ref{tab:smali}.

\begin{table}[ht]
  \centering
  \caption{An Exemplar Smali File}
  \label{tab:smali}
  \begin{tabular}{cccl}
    \toprule 
    \multicolumn{4}{c}{\textbf{Lbocb/lj/korsy/A;--\textgreater getIncomingSMS}} \\
    \midrule
    4 & 112 & invoke-direct & Ljava/lang/StringBuilder;--\textgreater \textless init\textgreater \\
    10 & 110 & invoke-virtual & Landroid/content/Intent;--\textgreater getAction \\
    22 & 110 & invoke-virtual & Ljava/lang/String;--\textgreater equals \\
    30 & 56 & if-eqz & [160] \\
    34 & 110 & invoke-virtual & Landroid/content/Intent;--\textgreater getExtras \\
    42 & 56 & if-eqz & [160] \\
    50 & 110 & invoke-virtual & Landroid/os/Bundle;--\textgreater get \\
    74 & 53 & if-ge & [160] \\
    90 & 113 & invoke-direct & Landroid/telephony/SmsMessage;--\textgreater  \\
     &  &  & createFromPdu \\
    98 & 110 & invoke-virtual & Landroid/telephony/SmsMessage;--\textgreater \\
     &  &  &  getDisplayOriginatingAddress \\
    106 & 110 & invoke-virtual & Ljava/lang/StringBuilder;--\textgreater append \\
    118 & 110 & invoke-virtual & Ljava/lang/StringBuilder;--\textgreater append \\
    126 & 110 & invoke-virtual & Landroid/telephony/SmsMessage;--\textgreater \\
     &  &  & getDisplayMessageBody \\
    134 & 110 & invoke-virtual & Ljava/lang/StringBuilder;--\textgreater append \\
    146 & 110 & invoke-virtual & Ljava/lang/StringBuilder;--\textgreater append \\
    156 & 40 & goto & [72] \\
    160 & 110 & invoke-virtual & Ljava/lang/StringBuilder;--\textgreater toString \\
    168 & 17 & return-object &  \\
    \midrule
    \multicolumn{4}{c}{\textbf{Lbocb/lj/korsy/A;--\textgreater onReceive}} \\
    \midrule
    8 & 112 & invoke-direct & Ljava/util/ArrayList;--\textgreater \textless init\textgreater \\
    14 & 110 & invoke-virtual & Lbocb/lj/korsy/A;--\textgreater getIncomingSMS \\
    22 & 114 & invoke-interface & Ljava/util/List;--\textgreater add \\
    32 & 110 & invoke-virtual & Ljava/lang/Object;--\textgreater toString \\
    40 & 110 & invoke-virtual & Lbocb/lj/korsy/A;--\textgreater sendTextMessage \\
    46 & 14 & return-void &  \\
    \midrule
    \multicolumn{4}{c}{\textbf{Lbocb/lj/korsy/A;--\textgreater sendTextMessage}} \\
    \midrule
    2 & 113 & invoke-static & Landroid/telephony/SmsManager;--\textgreater \\
     &  &  & getDefault \\
    18 & 116 & invoke-virtual & Landroid/telephony/SmsManager;--\textgreater \\
     &  & /range & sendTextMessage \\
    24 & 14 & return-void & \\
    \bottomrule
  \end{tabular}
\end{table}

\subsection{GAM and GAT Training}
\label{sec:gam_training}

Our primary goal was to localize malicious code within graphs. To achieve this, we required an explainable AI technique, as interpretability is crucial for understanding and verifying the decisions made by our model. Various explainable AI techniques exist, such as attention mechanisms, feature attribution, and model distillation, among others. Attention mechanisms allow the model to focus on the most relevant parts of the input data, enhancing interpretability by highlighting the critical components that influence the model's predictions. This is particularly beneficial in our context, where identifying specific portions of code that contribute to malicious behavior is essential.

We focused on graph machine learning techniques that incorporate attention mechanisms, specifically Graph Attention Mechanism (GAM) ~\cite{GAM_paper} and Graph Attention Network (GAT) ~\cite{velickovic2017gat, brody2021gat2}. GAM is particularly suited for malicious code localization as it emphasizes subgraph attention, identifying significant portions of the graph. GAT, in contrast, assigns attention to edges, weighing the importance of relationships between nodes, which helps capture the relevant interactions within the graph.

By combining GAM’s subgraph-level focus and GAT’s edge-level attention, we enhanced the precision of malicious code localization. This dual approach ensures a comprehensive and interpretable model, with transparent reasoning behind its decisions.

\subsubsection{Graph Attention Mechanism (GAM)}

The Graph Attention Model (GAM) is a neural network architecture designed to enhance graph classification tasks by focusing on the most informative parts of a graph. Below, we outline its layers and mathematical formulations.

\textbf{Input Layer: } The input graph is represented as \( G = (V, E) \), where \( V \) is the set of nodes and \( E \) is the set of edges. Each node \( v \in V \) is associated with a feature vector \( \mathbf{h}_v^{(0)} \in \mathbb{R}^d \), where \( d \) is the feature dimensionality.

\textbf{Structural Attention Layer:} This layer assigns an attention score to each node and selects the most informative subset of nodes $V_s \subseteq V$.

\paragraph{Node Importance Scoring}
The attention score for a node \( v \), denoted as \( A(v) \), is computed as:
\begin{equation}
A(v) = \text{softmax} \left( \mathbf{a}^\top \phi(\mathbf{h}_v^{(t)}, \mathbf{h}_\text{context}) \right),
\end{equation}
where \( \mathbf{a} \) is the learnable weight vector, \( \phi(\cdot) \) is the compatibility function (e.g., concatenation of node and context embeddings followed by a linear transformation), and \( \mathbf{h}_\text{context} \) is the current context vector summarizing the graph so far.

\paragraph{Node Selection}
Nodes with the highest scores \( A(v) \) are selected for further processing.

\textbf{Recurrent Neural Network (RNN) Layer:} The selected nodes are processed sequentially using a Long Short-Term Memory (LSTM) network.

\paragraph{State Update}
For each selected node $v$ at step $t$, the RNN updates its state:
\begin{equation}
\mathbf{h}^{(t)} = \text{RNN}(\mathbf{h}^{(t-1)}, \mathbf{h}_v^{(t)}),
\end{equation}
where $\mathbf{h}^{(t)}$ is the hidden state of the RNN at step $t$, and $\mathbf{h}_v^{(t)}$ is the feature vector of the selected node.

\textbf{External Memory Module: }The external memory aggregates information from selected nodes and maintains a global understanding of the graph.

\paragraph{Memory Update}
At each step $t$, the memory $\mathbf{M}$ is updated as:
\begin{equation}
\mathbf{M}^{(t)} = f_\text{update}(\mathbf{M}^{(t-1)}, \mathbf{h}^{(t)}),
\end{equation}
where $f_\text{update}$ is a learnable function.

\textbf{Output Layer:} The final graph embedding $\mathbf{z}_G$ is computed as:
\begin{equation}
\mathbf{z}_G = \text{Aggregate}(\{\mathbf{M}, \mathbf{h}^{(t)}\}),
\end{equation}
where $\text{Aggregate}$ is a pooling operation (e.g., mean, max, or concatenation).

For classification tasks, $\mathbf{z}_G$ is passed through a fully connected layer followed by softmax activation:
\begin{equation}
\hat{y} = \text{softmax}(\mathbf{W} \cdot \mathbf{z}_G + \mathbf{b}),
\end{equation}
where $\mathbf{W}$ and $\mathbf{b}$ are learnable parameters.

\subsubsection{Adaptation of GAM for MCL}

In this study, we apply the Graph Attention Mechanism (GAM) to the problem of malicious code localization, specifically targeting the binary graph classification of applications as either malicious or benign. GAM is used to identify the specific parts of code that distinguish malicious applications from benign software by assigning attention scores to each node. We utilized the existing GAM implementation ~\cite{gam_repo} available on GitHub.

In this study, the GAM is configured with a step size of 40, 10 agents, 50 epochs, and a learning rate of 0.0001. The step size is determined by an analysis of the API graphs. Although 688 APIs were included in the setup, not all graphs contained every API. With an average of 84 nodes per graph, the step size of 40 allows each agent to explore a substantial portion of the graph, covering approximately half of the average nodes. This balance ensures effective exploration while managing computational resources efficiently. The learning rate is selected according to the original study, while the number of agents and epochs are experimentally determined to balance effectiveness and running time. 

\subsubsection{Graph Attention Network (GAT)}

Graph Attention Networks (GAT) are a type of neural network designed to process graph-structured data by incorporating attention mechanisms. They are particularly effective for tasks such as node classification, link prediction, and graph classification.

Graphs represent data as nodes (entities) and edges (relationships). Traditional graph neural networks (GNNs) often assume that all neighbors contribute equally to a node's representation. GATs address this limitation by using attention mechanisms to assign different weights to different edges, allowing the model to focus on the most important neighbors.

The key components of a GAT are described below.

\textbf{Input:} The input to a GAT layer consists of a graph \( G = (V, E) \), where \( V \) is the set of nodes and \( E \) is the set of edges, and a node feature matrix \( \mathbf{X} \in \mathbb{R}^{N \times F} \), where \( N \) is the number of nodes and \( F \) is the number of features per node.

\textbf{Attention Mechanism:} The attention mechanism computes the importance of each neighbor for a given node. The steps are as follows:

\textbf{Feature Transformation:} Each node's features are transformed using a learnable weight matrix \( \mathbf{W} \), such that \( \mathbf{h}'_i = \mathbf{W} \mathbf{h}_i \), where \( \mathbf{h}_i \) is the feature vector of node \( i \).

\textbf{Attention Coefficients:} For a node $i$, the attention score with its neighbor $j$ is computed as:
\begin{equation}
\mathbf{e}_{ij} = \text{LeakyReLU}\left(\mathbf{a}^\top [\mathbf{h}'_i \| \mathbf{h}'_j]\right),
\end{equation}
where $\mathbf{a}$ is a learnable attention vector, $\|$ denotes vector concatenation, and $\mathbf{e}_{ij}$ is the unnormalized attention coefficient.

\textbf{Normalization:} The coefficients $\mathbf{e}_{ij}$ are normalized across all neighbors of node $i$ using the softmax function:
\begin{equation}
\alpha_{ij} = \frac{\exp(\mathbf{e}_{ij})}{\sum_{k \in \mathcal{N}(i)} \exp(\mathbf{e}_{ik})},
\end{equation}
where $\mathcal{N}(i)$ represents the set of neighbors of node $i$.

\textbf{Message Aggregation:} Each node aggregates information from its neighbors using the normalized attention coefficients:
\begin{equation}
\mathbf{h}''_i = \sigma\left(\sum_{j \in \mathcal{N}(i)} \alpha_{ij} \mathbf{h}'_j\right),
\end{equation}
where $\sigma$ is an activation function (e.g., ReLU).

\textbf{Multi-Head Attention:} To enhance expressiveness, GAT employs multiple attention heads. For $K$ attention heads, the outputs are either:
\begin{itemize}
    \item Concatenated (for intermediate layers):
    \begin{equation}
    \mathbf{h}''_i = \big\|_{k=1}^K \sigma\left(\sum_{j \in \mathcal{N}(i)} \alpha_{ij}^{(k)} \mathbf{h}'_j\right),
    \end{equation}
    \item Averaged (for the final layer):
    \begin{equation}
    \mathbf{h}''_i = \frac{1}{K} \sum_{k=1}^K \sigma\left(\sum_{j \in \mathcal{N}(i)} \alpha_{ij}^{(k)} \mathbf{h}'_j\right).
    \end{equation}
\end{itemize}

\subsubsection{Adaptation of GAT for MCL}

In this study, the Graph Attention Network (GAT) is configured with one input feature per node and eight hidden units in each layer. The model uses eight attention heads and predicts two output classes, distinguishing between malicious and benign nodes.

The training is carried out over 200 epochs with a learning rate of 0.0005. The network architecture includes multi-head GAT convolution, Exponential Linear Unit (ELU) activation, mean pooling for graph-level representation, and a final linear transformation layer.

Node attention scores, denoted as \( A(i) \), are computed by averaging the edge attention scores of neighboring nodes. This is represented mathematically as:
\begin{equation}
A(i) = \frac{1}{|\mathcal{N}(i)|} \sum_{j \in \mathcal{N}(i)} \alpha_{ij},
\end{equation}
where \( \mathcal{N}(i) \) is the set of neighbors of node \( i \), and \( \alpha_{ij} \) is the attention weight for the edge between nodes \( i \) and \( j \).

\subsection{Malicious Code Localization (MCL)}

After training our model with both GAM and GAT, attention values, denoted as $A_X(i)$, where $X \in \{\text{GAM}, \text{GAT}\}$, are assigned to the API calls made within each APK. 
These attention values play a crucial role in identifying malicious code. For each model, the method attention value for a method $m$ is given by:
\begin{equation}
MA_X(m) = \sum_{i \in \text{API\_calls}(m)} A_X(i).
\end{equation}

To ensure comparability, the method attention values are normalized so that the total attention values across all methods in an APK sum to 1:
\begin{equation}
MA_{X, \text{norm}}(m) = \frac{MA_X(m)}{\sum_{m' \in \text{Methods}} MA_X(m')}.
\end{equation}

The class attention value for a class $c$ is computed similarly:
\begin{equation}
CA_X(c) = \sum_{m \in \text{Methods}(c)} MA_{X, \text{norm}}(m).
\end{equation}

To classify methods and classes, an Attention Threshold-Based Selection approach is applied. Methods/classes are labeled \textbf{Malicious} if their normalized attention value \( A_{X, \text{norm}} \) exceeds a predefined threshold, and \textbf{Benign} otherwise.

For the final classification, the results from both GAM and GAT are combined using a logical AND operation. A method or class is classified as \textbf{Malicious} only if it is classified as malicious by both models; otherwise, it is classified as \textbf{Benign}.

\section{Experimental Results}
\label{sec:experimantal_results}

\subsection{Malware Detection}

The results in Table \ref{tab:malware_detection} compare our model’s performance under three configurations: using only GAM, only GAT, and a combination of both. This ablation study evaluates accuracy, precision, recall, F1-score, and FPR (False Positive Rate).

Using only GAM, the model achieves a high recall of 99.35\%, effectively identifying malware but at the cost of a high FPR (25.44\%), leading to 88.01\% accuracy and a 90.00\% F1-score. This suggests GAM captures true positives well but is prone to false positives. The GAT configuration improves accuracy (94.52\%) and significantly lowers FPR (8.57\%). It achieves 93.08\% precision and a 95.06\% F1-score, demonstrating a more balanced performance while maintaining strong recall (97.12\%). Combining GAM and GAT yields the best results, with 95.27\% accuracy, 6.89\% FPR, and 2.92\% FNR. This approach leverages the strengths of both models, achieving 94.36\% precision and a 95.70\% F1-score, enhancing malware detection while minimizing errors.

Overall, the combined configuration outperforms the others, balancing high recall with fewer FPR, making it the most robust choice for real-world malware detection.

\begin{table*}
  \caption{XAIDroid Malware Detection Results}
  \label{tab:malware_detection}
  \centering
  \begin{tabular}{ccccccc}
    \toprule
    \textbf{Test} & \textbf{FPR} & \textbf{FNR} & \textbf{Accuracy} & \textbf{Precision} & \textbf{Recall} & \textbf{F1-Score} \\
    \midrule
    Only GAM & 25.44\% & 0.65\% & 88.01\% & 82.26\% & 99.35\% & 90.00\% \\
    Only GAT & 8.57\% & 2.88\% & 94.52\% & 93.08\% & 97.12\% & 95.06\% \\
    Ensemble of GAM and GAT & 6.89\% & 2.92\% & 95.27\% & 94.36\% & 97.08\% & 95.70\% \\
    \bottomrule
  \end{tabular}
\end{table*}

\subsection{Malicious Code Localization}

\begin{table*}[ht]
  \centering
  \caption{XAIDroid Malicious Code Localization Metrics in Method and Class-Level}
  \label{tab:mcl-all}
  \begin{tabular}{ccccc|cccc}
    \toprule & 
    \multicolumn{4}{c}{\textbf{Method-Level Metrics}} & \multicolumn{4}{c}{\textbf{Class-Level Metrics}} \\ 
    \midrule
    \textbf{Malware Variant} & \textbf{Accuracy} & \textbf{Precision} & \textbf{Recall} & \textbf{F1 Score} & \textbf{Accuracy} & \textbf{Precision} & \textbf{Recall} & \textbf{F1 Score} \\
    \midrule
    Trojan.AndroidOS.Mobtes.co	& 95.53\% & 86.42\% & 98.36\% & 91.33\% & 97.04\% & 92.07\% & 98.60\% & 94.55\% \\
    Trojan-SMS.AndroidOS.FakeInst.a	& 97.55\% & 81.37\% & 98.06\% & 87.07\% & 98.49\% & 93.78\% & 99.58\% & 96.26\% \\
    Trojan-SMS.AndroidOS.Agent.aax & 97.26\% & 90.16\% & 99.05\% & 91.98\% & 99.31\% & 100.00\% & 97.19\% & 98.46\% \\
    Trojan-SMS.AndroidOS.Opfake.bo & 97.01\% & 90.14\% & 98.55\% & 92.80\% & 96.39\% & 95.17\% & 100.00\% & 96.70\% \\
    Trojan-Dropper.AndroidOS.Wapnor.a & 100.00\% & 100.00\% & 100.00\% & 100.00\% & 100.00\% & 100.00\% & 100.00\% & 100.00\% \\
    Trojan-Spy.AndroidOS.Agent.bz & 99.15\% & 100.00\% & 92.80\% & 96.25\% & 98.47\% & 100.00\% & 93.20\% & 96.45\% \\
    Trojan-Banker.AndroidOS.Gepew.b	& 96.03\% & 94.79\% & 91.57\% & 93.03\% & 97.91\% & 100.00\% & 95.91\% & 97.91\% \\
    AdWare.AndroidOS.Apofer.b & 98.20\% & 90.45\% & 96.94\% & 93.31\% & 96.66\% & 91.24\% & 97.42\% & 93.51\% \\
    Trojan-Spy.AndroidOS.Agent.qu & 95.13\% & 84.07\% & 99.72\% & 90.50\% & 96.22\% & 81.26\% & 99.35\% & 87.91\% \\
    \midrule
    \textbf{Average} & \textbf{97.32\%} & \textbf{90.82\%} & \textbf{97.23\%} & \textbf{92.92\%} & \textbf{97.83\%} & \textbf{94.84\%} & \textbf{97.92\%} & \textbf{95.75\%} \\
    \bottomrule
  \end{tabular}
\end{table*}

\begin{figure*}[t]
    \centering
    \includegraphics[width=0.8\textwidth]{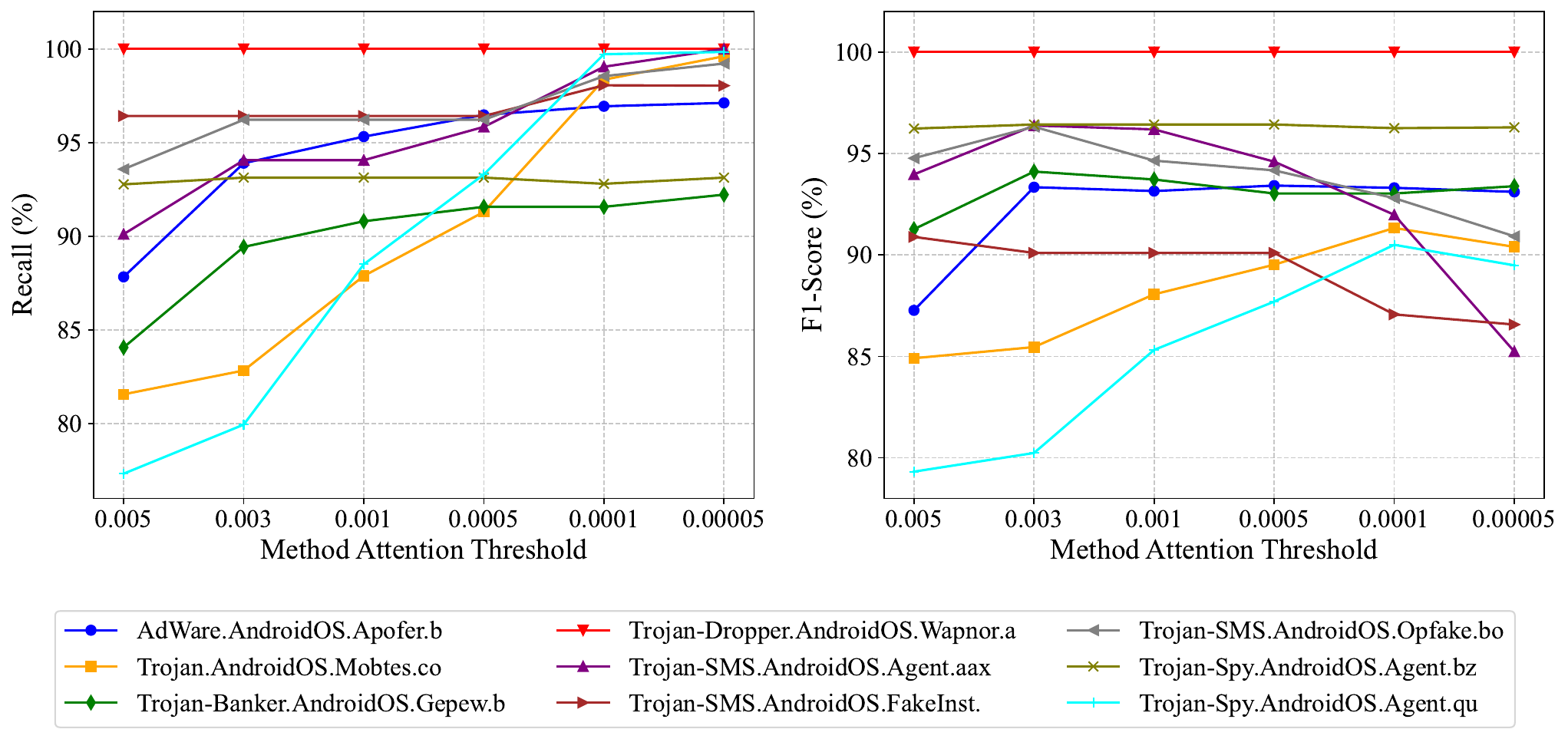}
    \caption{Method Level MCL - Recall and F1-Score Metrics}
    \label{fig:combined_metrics}
\end{figure*}

To evaluate the detection of malicious code snippets at the method or class level, we compared our model’s predictions against a baseline of known malicious methods and classes. The evaluation metrics include accuracy, precision, recall, and F1-score. \textit{Accuracy} measures the proportion of correctly identified methods and classes. \textit{Precision} reflects the proportion of predicted malicious methods/classes that are truly malicious. \textit{Recall} indicates the proportion of actual malicious methods/classes correctly identified. \textit{F1-score} balances precision and recall, considering both false positives and negatives.

The recall and F1-score metrics for method-level MCL across different malware variants, evaluated at various attention-threshold levels, are illustrated in Figure \ref{fig:combined_metrics}. The x-axis represents the attention threshold, decreasing from left to right. At a threshold of 0.005, recall rates range from 77\% to 100\%, increasing as the threshold lowers. When set to 0.0001, the average recall rate reaches 97.23\%, with an average F1-score of 92.92\%. Lowering the threshold further to 0.00005 yields a similar recall rate but reduces the F1-score to 92.00\%, making 0.0001 the optimal threshold for this setup. Method-level MCL metrics per malware variant (threshold 0.0001) and class-level MCL metrics (threshold 0.001) are detailed in Table \ref{tab:mcl-all} and further analyzed in the following subsections.

\subsubsection{Trojan.AndroidOS.Mobtes.co}

This is a general-purpose trojan capable of executing various malicious activities on an infected device, including stealing sensitive data, sending premium-rate SMS messages, and downloading additional malware. The "Mobtes" family is particularly known for spying on users by intercepting SMS messages and calls. In this variant, API calls related to phone state listeners, HTTP posting, SMS sending, and string manipulations receive higher attention values. As a result, XAIDroid accurately identifies malicious methods, such as transmitting call details via HTTP or sending sensitive data via SMS, achieving a recall rate of 98.36\%. Analyzing missed malicious methods shows they are linked to identified ones—either calling or being called by them—making them easily detectable for security analysts using XAIDroid’s outputs.

\subsubsection{Trojan-SMS.AndroidOS.FakeInst.a}

This malware disguises itself as a legitimate application, often mimicking popular games or utilities to deceive users into installing it. Once installed, it covertly sends SMS messages to premium-rate numbers, resulting in unexpected charges. XAIDroid assigns high attention values to key API calls, including append and init (StringBuilder) and sendTextMessage and getDefault (SmsManager). Leveraging these indicators, it accurately identifies malicious methods—such as those responsible for sending messages or initiating these processes (e.g., start or onReceive)—achieving a recall rate of 98.06\%. An analysis of missed malicious methods (falsely labeled as benign) reveals that they contain very few API calls, leading to lower attention values despite embedding high-attention APIs. As a result, these methods are overlooked; however, their invoking methods are flagged as malicious, assisting security analysts in pinpointing them more effectively.

\subsubsection{Trojan-SMS.AndroidOS.Agent.aax}

This variant exploits SMS functionalities, typically targeting the interception or sending of SMS messages. Its primary goal is often to steal authentication codes (e.g., two-factor authentication codes) or to enroll the user in unwanted services. XAIDroid assigns high attention to Android API calls such as sendTextMessage (SmsManager) and execute (HttpClient), recognizing their frequent use in malicious activities associated with this variant. Our methodology accurately identifies malicious methods within APKs, achieving a recall rate of 99.05\%. These methods include reading sensitive data from Android’s SharedPreferences, transmitting information via SMS, extracting device identifiers, and registering devices with remote servers, all of which highlight privacy and security risks. XAIDroid successfully detects all malicious methods in 14 out of 15 APKs. The missed methods in the remaining APK involve initiating broadcast receivers and scheduling alarms to support malicious activities.

\subsubsection{Trojan-SMS.AndroidOS.Opfake.bo}

This variant disguises itself as popular apps or websites to deceive users, sending SMS messages to premium numbers, subscribing users to costly services, or harvesting personal information. XAIDroid detects malicious methods, such as retrieving values from SharedPreferences, generating unique installation IDs, processing sensitive configurations using string manipulations, and storing or transmitting them over HTTP, achieving a recall rate of 98.55\%. All malicious methods are identified in 13 out of 15 APKs, with one APK missing 1 out of 10 methods and another missing 2 out of 17. The missed methods contain only a few API calls, resulting in lower attention values, but since they are invoked within correctly identified malicious methods, security analysts can easily detect them.

\subsubsection{Trojan-Dropper.AndroidOS.Wapnor.a}

A trojan dropper’s primary function is to install additional malicious components or payloads onto an infected device. The \textit{Wapnor} variant might drop additional malware that can steal information, control the device remotely, or perform other harmful actions. Typically, the dropper conceals its activities, making detection more challenging. XAIDroid successfully identified malicious methods in this variant, including those that dynamically manipulate classes and fields, load additional code, or extract resources from the application's assets and internal storage, achieving  100.00\% recall rate.

\subsubsection{Trojan-Spy.AndroidOS.Agent.bz}

This spyware variant monitors user activities and steals sensitive data, such as text messages, call logs, and location, which is sent to the attacker’s server for malicious purposes like identity theft or blackmail. XAIDroid flags Android API calls related to HttpClient execution and FileOutputStream operations with high attention values, correctly identifying methods responsible for accessing storage and performing file uploads via HTTP, achieving a 92.80\% recall rate. However, access to the phone number (via getLine1Number) and URL manipulation (for redirection to untrusted sources) are not flagged due to their isolation in the code graph, as these methods are not directly invoked by other functions and remain undetected. It’s possible these code blocks are triggered by a malicious payload, downloaded or executed at runtime. This is supported by Java reflection APIs, network activity suggesting remote code fetching, and the use of the JavaScript interface to fetch and execute additional payloads, all pointing to dynamic code loading.

\subsubsection{Trojan-Banker.AndroidOS.Gepew.b}

Banking trojans are designed to steal financial information, such as online banking credentials or credit card details. The \textit{Gepew} variant may overlay fake login pages over legitimate banking apps or websites to deceive users into entering their details, which are then sent to the attacker. In this variant, XAIDroid detects malicious methods that establish HTTP connections, send unauthorized data (such as user details or contacts) to a remote server, and download files from remote servers to the device’s storage, potentially enabling the installation of harmful or unauthorized content, with a recall rate of 91.57\%. These malicious methods facilitate the downloading of files that may contain code designed to capture sensitive banking information. By manipulating network connections and files, the app can secretly exfiltrate financial data or exploit vulnerabilities to steal banking credentials.

\subsubsection{AdWare.AndroidOS.Apofer.b}

This is a form of adware that aggressively pushes advertisements onto the infected device. While it may not be as harmful as other types of malware, it can still be very disruptive, slowing down the device, draining battery life, and potentially leading to unwanted downloads or installations. In this variant, API calls related to the execution of HttpClient, the getSharedPreferences method of Context, and the append method of StringBuilder receive higher attention values. These methods are associated with reading device-related data, which may include authentication tokens, user identifiers, application settings, and other private app data. Additionally, HTTP requests and URL opening via the browser are flagged as potentially malicious. These API calls are utilized to show advertisement websites based on the user's information, enabling targeted ad behavior and potentially intrusive actions. XAIDroid achieves a recall rate of 96.94\% in this variant.

\subsubsection{Trojan-Spy.AndroidOS.Agent.qu}

This variant is a form of spyware similar to the \textit{Agent.bz} variant, capable of stealing a wide range of data, including SMS messages, call logs, browser history, and keystrokes. Operating silently in the background, it often remains undetected by the user, posing significant privacy risks. This variant accurately identifies all malicious methods in 75 out of 77 APK samples. Android API calls related to SMS sending and string manipulation are given higher attention, ensuring precise detection of methods for accessing contact lists, reading inbox messages, and sending multipart SMS with a recall rate of 99.72\%. In the remaining two APKs, one method is missed in each, out of 11 and 8 malicious methods, respectively. The missed methods, involving access to IMEI and IMSI numbers, receive lower attention but can still be identified by security analysts, as they are located within classes flagged as malicious by XAIDroid.

In summary, XAIDroid exhibits strong performance in malicious code localization (MCL), achieving high recall rates and F1-scores. The few missed methods are either related to detected malicious methods, making them easily identifiable by a security analyst, or exist in isolated code paths that are challenging to detect. The latter indicates that these methods are likely activated through dynamic code loading, enabling them to bypass traditional static analysis.

\subsection{Comparison with Other Studies}

\begin{table*}[ht]
  \centering
  \caption{Comparison with MKLDroid \cite{MKLDROID_paper} and DexBERT \cite{dexbert}}
  \label{tab:mkldroid_comparison}
  \begin{tabular}{cccccccc}
    \toprule
     & Level & Methodology & FPR & FNR & Precision & Recall & F1-Score \\
    \midrule
    MKLDroid ~\cite{MKLDROID_paper} & Class & Top-10 Classes & 17.00\% & 5.67\% & 14.34\% & 94.33\% & 24.88\% \\
    DexBERT \cite{dexbert} & Class & Class-level classification & 0.06\% & 0.21\% & 99.83\% & 99.79\% & 99.81\% \\
    XAIDroid & Class & Attention-threshold: 0.001 & 3.42\% & 1.33\% & 91.12\% & 98.67\% & 93.97\% \\
    XAIDroid & Method & Attention-threshold: 0.0001 & 5.31\% & 1.52\% & 86.22\% & 98.48\% & 91.26\% \\
    \bottomrule
  \end{tabular}
\end{table*}

MKLDroid \cite{MKLDROID_paper} and DexBERT \cite{dexbert} also focus on localizing malicious code and present their results using the Mystique \cite{MYSTIQUE_paper} dataset at the class level. In this section, we will compare our results with those from these studies.

MKLDroid \cite{MKLDROID_paper} generates five different graphs for each APK file, trains them using Multiple Kernel Training, and assigns maliciousness scores (referred to as m-scores in their terminology) to each class within the APKs. Then it identifies the top 10 classes with the highest m-scores as malicious. On the other hand, the DexBERT study \cite{dexbert} creates sequence embeddings for each class in an APK file individually, classifying them as benign or malicious.

We compared our malicious code localization at the class and method level with MKLDroid \cite{MKLDROID_paper} and DexBERT \cite{dexbert}. Since their work used only the Mystique dataset, we present our class- and method-level classification results exclusively for Mystique \cite{MYSTIQUE_paper} in this table to ensure a fair comparison. Two of the nine variants used in our research, specifically Trojan.AndroidOS.Mobtes.co and Trojan-Spy.AndroidOS.Agent.qu, are from the Mystique dataset, and we utilized them in this comparison.

As shown in the table, we outperformed MKLDroid in all metrics in class-level MCL. When comparing our class-level recall rates with DexBERT, we observed similar performance, achieving a recall rate of 98.67\%, compared to DexBERT’s 99.79\%. However, our class-level F1 score (93.97\%) is lower than DexBERT’s 99.81\%.

DexBERT ~\cite{dexbert} extracts all class files from both benign and malware training APKs, assigns them labels as either benign or malicious, and fine-tunes their BERT classifier using these class labels. This method has two key limitations. First, DexBERT ~\cite{dexbert} primarily relies on the Mystique ~\cite{MYSTIQUE_paper} dataset, which consists of artificially generated malware with a highly similar code base. As a result, the model may benefit from these inherent similarities rather than learning to generalize to diverse, real-world malware, potentially resulting in overfitting. Comprehensive evaluation on a broader set of real malware samples is necessary to assess the robustness of the proposed approach. Second, in practical malware detection scenarios, predefined datasets with labeled individual classes do not exist. Creating such a dataset would require extensive manual effort by security experts, making DexBERT’s ~\cite{dexbert} training approach difficult to scale for real-world applications.

In contrast, our proposed methodology XAIDroid does not rely on pre-labeled class-level data. Instead, it leverages API call graphs extracted from both benign and malware samples, employing explainable AI techniques to distinguish between them. XAIDroid autonomously identifies malicious code segments without requiring predefined class-level labels. Moreover, XAIDroid supports malicious code localization at both the class and method levels — a more fine-grained and challenging task. This flexibility, combined with high performance on both artificial and real malware datasets, highlights the effectiveness and practical applicability of the proposed approach. 

\section{Threats to Validity}
\label{sec:threats_to_validity}

While our proposed method demonstrates promising results in localizing Android malicious code snippets by using GAM and GAT, several potential limitations and threats to validity should be acknowledged.

\textbf{Challenges in Establishing a Malicious Code Localization Baseline:} One major challenge we encountered was the absence of a public MCL baseline for researchers. During our comparison of MCL results, we found no universally accepted or publicly available baseline. Although we initially obtained the Mystique \cite{MYSTIQUE_paper} baseline from MKLDroid \cite{MKLDROID_paper}, it only provided class-level MCL data, which required us to develop our own method-level baseline. Throughout this process, we identified and corrected several errors in the class-level MCL data to ensure the accuracy of our comparisons. Furthermore, we sought to acquire 100 manually-analyzed APKs from the AMD \cite{AMD} dataset via Droidetec \cite{DROIDETEC_paper}, but our request went unanswered, complicating our efforts. Consequently, we had to create our own MCL baseline, which was both time-consuming and resource-intensive. Although we are sharing this baseline on our GitHub account for other researchers, it remains limited in scope. Comprehensive MCL baselines from antivirus firms, leveraging their resources and expertise, would accelerate research.

\textbf{Static Analysis Limitations:} The proposed approach relies on static analysis to extract API call graphs, which may overlook malware behaviors involving dynamic code loading or reflection at runtime. While static analysis can capture API calls related to dynamic code loading, it cannot access the content of dynamically loaded code segments executed at runtime. To address this limitation, integrating dynamic analysis or adopting hybrid models that combine static and dynamic techniques could improve detection capabilities, particularly for malware with sophisticated runtime behaviors.

\textbf{Adversarial Attacks on Graphs:} Although our model demonstrates high recall rates on datasets like Mystique \cite{MYSTIQUE_paper}, AMD \cite{AMD} and CICMalDroid \cite{maldroid_1, maldroid_2}, there is concern about its susceptibility to adversarial attacks \cite{sun2022adversarial}. Malware authors might manipulate the API call graph structure or embed features in ways that evade detection. Potential adversarial strategies could include altering the graph topology, injecting noise into the embeddings, or exploiting vulnerabilities in the learning process. While representing applications as API call graphs provides resilience against certain obfuscation techniques, such as renaming, the model remains potentially vulnerable to control flow obfuscation. This aspect requires further investigation.

\section{Conclusion}
\label{sec:conclusion}

XAIDroid introduces a novel method for localizing malicious code within Android applications by representing them as API call graphs, which inherently preserve code semantics. Subsequently, the problem is framed as a graph classification task. A key innovation of this research lies in the integration of both subgraph-level and edge-level attention mechanisms for the first time to address the challenge of finding malicious code as a whole. 

The proposed approach is evaluated at two levels: class level and method level, for coarse- and fine-grained localization. The evaluation is carried out on three datasets. The first data set called Mystique \cite{MYSTIQUE_paper} is synthetically generated using features of existing malware and evolutionary computation techniques. The second and third data sets are a subset of a well-accepted dataset, AMD \cite{AMD} and CICMalDroid 2020 ~\cite{maldroid_1, maldroid_2}, in the research community. Furthermore, to assess the performance of the proposed approach in real malware, 105 samples of AMD \cite{AMD} and CICMalDroid 2020 ~\cite{maldroid_1, maldroid_2} are manually analyzed to tag their malicious code snippets. The experimental results show a higher recall rate for each data set at each level while maintaining a good F1 score value.

Although MCL research is less common than broader malware detection, studies such as MKLDroid \cite{MKLDROID_paper} and DexBERT \cite{dexbert} primarily focus on class-level MCL, relying on artificially generated data sets such as Mystique \cite{MYSTIQUE_paper}. In contrast, our work advances the field by incorporating real-world malware samples into training and evaluation, ensuring a more robust, generalizable approach. Moreover, unlike previous studies limited to class-level localization, we achieve method-level localization. This finer granularity enhances the identification of malicious codes and fills a critical research gap, setting a new benchmark for MCL. Furthermore, our method eliminates the need for prelabeled malicious or benign classes, as required in DexBERT \cite{dexbert}, making it more practical for real-world scenarios. 

In conclusion, automating malicious code localization is a pivotal advancement in cybersecurity. With rising demand for malware analysts and increasing unknown malware, automation is essential. Fine-grained localization can significantly reduce the time of manual analysis, enhance scalability, and enable novel malware detection approaches focused on critical code segments. Moreover, these advances support explainable AI by providing transparent insights into the detection process, fostering a clearer understanding of flagged malicious code.

\section{Ethical Considerations}
\label{sec:ethical}

In conducting this research on malware detection and localization within Android APKs, careful attention has been given to the ethical implications, particularly the potential mislabeling of legitimate third-party libraries as malicious. Recognizing that incorrect labeling could lead to significant negative outcomes, such as financial losses for developers or companies, we have implemented several precautions to minimize these risks.

\textbf{Efforts to Avoid Mislabeling:} During the manual localization of malicious classes and methods, we made a concerted effort to differentiate between genuinely malicious code and legitimate code used by third-party libraries. This process involved rigorous analysis and cross-referencing with known benign libraries to ensure that they were not wrongly classified as malicious. The methodology used was designed to be as accurate as possible to prevent harm to legitimate entities.

\textbf{Ongoing Ethical Responsibility:} We understand the serious implications that our research findings can have on individuals, companies, and the broader community. Therefore, we will continue to monitor and address any ethical issues that arise after publication. This commitment to ethical responsibility extends beyond the initial research and publication phase, ensuring that any unintended negative consequences are quickly and appropriately managed.

By implementing these safeguards, we aim to ensure ethical research while contributing valuable insights to malware analysis.

\section*{Acknowledgment}

Merve Cigdem Ipek thanks Aselsan Inc. for their support during PhD studies.

\ifCLASSOPTIONcaptionsoff
  \newpage
\fi

\bibliographystyle{IEEEtran}
\bibliography{bibtex/references.bib}

\begin{thebibliography}{10}
\providecommand{\url}[1]{#1}
\csname url@samestyle\endcsname
\providecommand{\newblock}{\relax}
\providecommand{\bibinfo}[2]{#2}
\providecommand{\BIBentrySTDinterwordspacing}{\spaceskip=0pt\relax}
\providecommand{\BIBentryALTinterwordstretchfactor}{4}
\providecommand{\BIBentryALTinterwordspacing}{\spaceskip=\fontdimen2\font plus
\BIBentryALTinterwordstretchfactor\fontdimen3\font minus \fontdimen4\font\relax}
\providecommand{\BIBforeignlanguage}[2]{{%
\expandafter\ifx\csname l@#1\endcsname\relax
\typeout{** WARNING: IEEEtran.bst: No hyphenation pattern has been}%
\typeout{** loaded for the language `#1'. Using the pattern for}%
\typeout{** the default language instead.}%
\else
\language=\csname l@#1\endcsname
\fi
#2}}
\providecommand{\BIBdecl}{\relax}
\BIBdecl

\bibitem{trendmicro2021}
T.~Micro. (2021) Trend micro cloud app security threat report 2021. \url{https://www.trendmicro.com/vinfo/ru/security/research-and-analysis/threat-reports/roundup/trend-micro-cloud-app-security-threat-report-2021}.

\bibitem{trendmicro2024}
------. (2024) Pushing the outer limits: Trend micro 2024 midyear cybersecurity threat report. \url{https://www.trendmicro.com/vinfo/us/security/research-and-analysis/threat-reports/roundup/pushing-the-outer-limits-trend-micro-2024-midyear-cybersecurity-threat-report}.

\bibitem{MSDroid}
Y.~He, Y.~Liu, L.~Wu, Z.~Yang, K.~Ren, and Z.~Qin, ``Msdroid: Identifying malicious snippets for android malware detection,'' \emph{IEEE Transactions on Dependable and Secure Computing}, vol.~20, no.~3, pp. 2025--2039, 2023.

\bibitem{gray2021identifying}
J.~Gray, D.~Sgandurra, and L.~Cavallaro, ``Identifying authorship style in malicious binaries: techniques, challenges \& datasets,'' \emph{arXiv preprint arXiv:2101.06124}, 2021.

\bibitem{huang2018large}
H.~Huang, C.~Zheng, J.~Zeng, W.~Zhou, S.~Zhu, P.~Liu, I.~Molloy, S.~Chari, C.~Zhang, and Q.~Guan, ``A large-scale study of android malware development phenomenon on public malware submission and scanning platform,'' \emph{IEEE Transactions on Big Data}, vol.~7, no.~2, pp. 255--270, 2021.

\bibitem{cozzi2020tangled}
E.~Cozzi, P.-A. Vervier, M.~Dell'Amico, Y.~Shen, L.~Bilge, and D.~Balzarotti, ``The tangled genealogy of iot malware,'' in \emph{Proceedings of the 36th Annual Computer Security Applications Conference}, 2020, pp. 1--16.

\bibitem{wu2020comprehensive}
Z.~Wu, S.~Pan, F.~Chen, G.~Long, C.~Zhang, and P.~S. Yu, ``A comprehensive survey on graph neural networks,'' \emph{IEEE Transactions on Neural Networks and Learning Systems}, vol.~32, no.~1, pp. 4--24, 2020.

\bibitem{GAM_paper}
J.~B. Lee, R.~Rossi, and X.~Kong, ``Graph classification using structural attention,'' in \emph{Proceedings of the 24th ACM SIGKDD International Conference on Knowledge Discovery \& Data Mining}, 2018, pp. 1666--1674.

\bibitem{velickovic2017gat}
P.~Veličković, G.~Cucurull, A.~Casanova, A.~Romero, P.~Liò, and Y.~Bengio, ``Graph attention networks,'' \emph{6th International Conference on Learning Representations}, 2017.

\bibitem{brody2021gat2}
\BIBentryALTinterwordspacing
S.~Brody, U.~Alon, and E.~Yahav, ``How attentive are graph attention networks?'' \emph{CoRR}, vol. abs/2105.14491, 2021. [Online]. Available: \url{http://dblp.uni-trier.de/db/journals/corr/corr2105.html#abs-2105-14491}
\BIBentrySTDinterwordspacing

\bibitem{MYSTIQUE_paper}
G.~Meng, Y.~Xue, C.~Mahinthan, A.~Narayanan, Y.~Liu, J.~Zhang, and T.~Chen, ``Mystique: Evolving android malware for auditing anti-malware tools,'' in \emph{Proceedings of the 11th ACM on Asia conference on computer and communications security}, 2016, pp. 365--376.

\bibitem{AMD}
Y.~Li, J.~Jang, X.~Hu, and X.~Ou, ``Android malware clustering through malicious payload mining,'' in \emph{International Symposium on Research in Attacks, Intrusions, and Defenses. Springer, Cham}, 2017, pp. 192--214.

\bibitem{maldroid_1}
D.~S. Keyes, B.~Li, G.~Kaur, A.~H. Lashkari, F.~Gagnon, and F.~Massicotte, ``Entroplyzer: Android malware classification and characterization using entropy analysis of dynamic characteristics,'' \emph{Reconciling Data Analytics, Automation, Privacy, and Security: A Big Data Challenge (RDAAPS)}, 2021.

\bibitem{maldroid_2}
A.~Rahali, A.~H. Lashkari, G.~Kaur, L.~Taheri, F.~Gagnon, and F.~Massicotte, ``Didroid: Android malware classification and characterization using deep image learning,'' \emph{10th International Conference on Communication and Network Security (ICCNS2020)}, pp. 70--82, 2020.

\bibitem{googleplay}
Google. (2024) Google play store. \url{https://play.google.com/}, Accessed: 2024.

\bibitem{apkpure}
Apkpure. (2024) Apkpure. \url{https://apkpure.com/}, Accessed: 2024.

\bibitem{DROIDETEC_paper}
Z.~Ma, H.~Ge, Z.~Wang, Y.~Liu, and X.~Liu, ``Droidetec: Android malware detection and malicious code localization through deep learning,'' \emph{arXiv preprint arXiv:2002.03594}, 2020.

\bibitem{MKLDROID_paper}
A.~Narayanan, M.~Chandramohan, L.~Chen, and Y.~Liu, ``A multi-view context-aware approach to android malware detection and malicious code localization,'' \emph{Empirical Software Engineering}, vol.~23, pp. 1222--1274, 2018.

\bibitem{dexbert}
T.~Sun, K.~Allix, K.~Kim, X.~Zhou, D.~Kim, D.~Lo, T.~F. Bissyandé, and J.~Klein, ``Dexbert: Effective, task-agnostic and fine-grained representation learning of android bytecode,'' \emph{IEEE Transactions on Software Engineering}, vol.~49, no.~10, pp. 4691--4706, October 2023.

\bibitem{ma2019combination}
Z.~Ma, H.~Ge, Y.~Liu, M.~Zhao, and J.~Ma, ``A combination method for android malware detection based on control flow graphs and machine learning algorithms,'' \emph{IEEE access}, vol.~7, pp. 21\,235--21\,245, 2019.

\bibitem{pektacs2020deep}
A.~Pekta{\c{s}} and T.~Acarman, ``Deep learning for effective android malware detection using api call graph embeddings,'' \emph{Soft Computing}, vol.~24, pp. 1027--1043, 2020.

\bibitem{gao2021gdroid}
H.~Gao, S.~Cheng, and W.~Zhang, ``Gdroid: Android malware detection and classification with graph convolutional network,'' \emph{Computers \& Security}, vol. 106, p. 102264, 2021.

\bibitem{Amdase}
Y.~Hongyu, W.~Youwei, Z.~Liang, C.~Xiang, and Z.~Hu, ``A novel android malware detection method with api semantics extraction,'' \emph{Computers \& Security}, vol. 137, 2024.

\bibitem{li2022dmalnet}
C.~Li, Z.~Cheng, H.~Zhu, L.~Wang, Q.~Lv, Y.~Wang, N.~Li, and D.~Sun, ``Dmalnet: Dynamic malware analysis based on api feature engineering and graph learning,'' \emph{Computers \& Security}, vol. 122, p. 102872, 2022.

\bibitem{fu2022encrypted}
Z.~Fu, M.~Liu, Y.~Qin, J.~Zhang, Y.~Zou, Q.~Yin, Q.~Li, and H.~Duan, ``Encrypted malware traffic detection via graph-based network analysis,'' in \emph{Proceedings of the 25th International Symposium on Research in Attacks, Intrusions and Defenses}, 2022, pp. 495--509.

\bibitem{narayanan2016contextual}
A.~Narayanan, G.~Meng, L.~Yang, J.~Liu, and L.~Chen, ``Contextual weisfeiler-lehman graph kernel for malware detection,'' in \emph{2016 International Joint Conference on Neural Networks (IJCNN)}.\hskip 1em plus 0.5em minus 0.4em\relax IEEE, 2016, pp. 4701--4708.

\bibitem{SMO-MKL}
Z.~Sun, N.~Ampornpunt, M.~Varma, and S.~Vishwanathan, ``Multiple kernel learning and the smo algorithm,'' \emph{Advances in neural information processing systems}, vol.~23, 2010.

\bibitem{mcl_Android_GNN}
Q.~Wu, P.~Sun, X.~Hong, X.~Zhu, and B.~Liu, ``An android malware detection and malicious code location method based on graph neural network,'' in \emph{Proceedings of the 2021 4th International Conference on Machine Learning and Machine Intelligence}, 2021, pp. 50--56.

\bibitem{piggybacked_li_1}
L.~Li, D.~Li, T.~F. Bissyand{\'e}, J.~Klein, H.~Cai, D.~Lo, and Y.~Le~Traon, ``Automatically locating malicious packages in piggybacked android apps,'' in \emph{2017 IEEE/ACM 4th International Conference on Mobile Software Engineering and Systems (MOBILESoft)}.\hskip 1em plus 0.5em minus 0.4em\relax IEEE, 2017, pp. 170--174.

\bibitem{piggybacked_li_2}
------, ``On locating malicious code in piggybacked android apps,'' \emph{Journal of Computer Science and Technology}, vol.~32, pp. 1108--1124, 2017.

\bibitem{suarez2018eight}
G.~Suarez-Tangil and G.~Stringhini, ``Eight years of rider measurement in the android malware ecosystem: evolution and lessons learned,'' \emph{arXiv preprint arXiv:1801.08115}, 2018.

\bibitem{wang2018beyond}
H.~Wang, Z.~Liu, J.~Liang, N.~Vallina-Rodriguez, Y.~Guo, L.~Li, J.~Tapiador, J.~Cao, and G.~Xu, ``Beyond google play: A large-scale comparative study of chinese android app markets,'' in \emph{Proceedings of the Internet Measurement Conference 2018}, 2018, pp. 293--307.

\bibitem{wei2017deep}
F.~Wei, Y.~Li, S.~Roy, X.~Ou, and W.~Zhou, ``Deep ground truth analysis of current android malware,'' in \emph{Detection of Intrusions and Malware, and Vulnerability Assessment: 14th International Conference, DIMVA 2017, Bonn, Germany, July 6-7, 2017, Proceedings 14}.\hskip 1em plus 0.5em minus 0.4em\relax Springer, 2017, pp. 252--276.

\bibitem{salem2018idea}
A.~Salem, T.~Schmidt, and A.~Pretschner, ``Idea: Automatic localization of malicious behaviors in android malware with hidden markov models,'' in \emph{Engineering Secure Software and Systems: 10th International Symposium, ESSoS 2018, Paris, France, June 26-27, 2018, Proceedings 10}.\hskip 1em plus 0.5em minus 0.4em\relax Springer, 2018, pp. 108--115.

\bibitem{virus_total}
VirusTotal. (2024) Virustotal. \url{https://www.virustotal.com/}.

\bibitem{genome}
Y.~Zhou and X.~Jiang, ``Dissecting android malware: Characterization and evolution,'' in \emph{2012 IEEE Symposium on Security and Privacy}, 2012, pp. 95--109.

\bibitem{kaspersky}
K.~Lab. (2024) Kaspersky. \url{https://www.kaspersky.com/}.

\bibitem{koodous}
Koodous. (2024) Koodous. \url{https://www.koodous.com/}.

\bibitem{axplorer}
M.~Backes, S.~Bugiel, E.~Derr, P.~McDaniel, D.~Octeau, and S.~Weisgerber, ``On demystifying the android application framework: Re-visiting android permission specification analysis,'' in \emph{Proceedings of the 25th USENIX Security Symposium, August 10–12, 2016, Austin, TX}.\hskip 1em plus 0.5em minus 0.4em\relax USENIX Association, 2016, pp. 1101--1118.

\bibitem{aafer2018precise}
Y.~Aafer, G.~Tao, J.~Huang, X.~Zhang, and N.~Li, ``Precise android api protection mapping derivation and reasoning,'' in \emph{Proceedings of the 2018 ACM SIGSAC Conference on Computer and Communications Security}, 2018, pp. 1151--1164.

\bibitem{gam_repo}
B.~Rozemberczki, ``Gam,'' 2022, \url{https://github.com/benedekrozemberczki/GAM}.

\bibitem{sun2022adversarial}
L.~Sun, Y.~Dou, C.~Yang, K.~Zhang, J.~Wang, S.~Y. Philip, L.~He, and B.~Li, ``Adversarial attack and defense on graph data: A survey,'' \emph{IEEE Transactions on Knowledge and Data Engineering}, 2022.

\end{thebibliography}

\begin{IEEEbiography}[
{\includegraphics[width=1in,height=1.25in,clip,keepaspectratio]{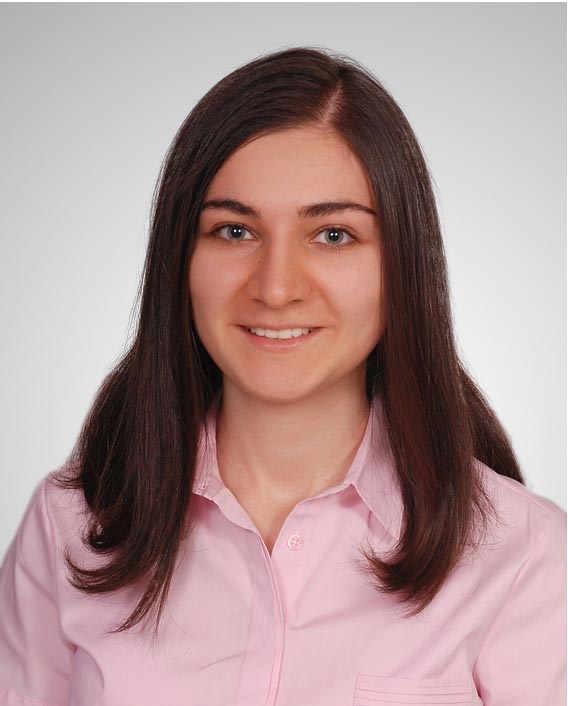}}
]
{Merve Cigdem Ipek} received the B.Sc. degree in Electrical and Electronics Engineering
from Middle East Technical University, Ankara, Turkey, in 2015, and
the M.Sc. degree from the Electrical and Electronics Engineering, Hacettepe University, Ankara, Turkey, in 2019. She is currently pursuing the Ph.D. degree in Computer Engineering, Hacettepe University, Ankara, Turkey. Her current
research interests include Android system security, artificial intelligence and machine learning.
\end{IEEEbiography}

\vskip 0pt plus -1fil
\begin{IEEEbiography}[
{\includegraphics[width=1.25in,height=1.5in,clip,keepaspectratio, angle=270]{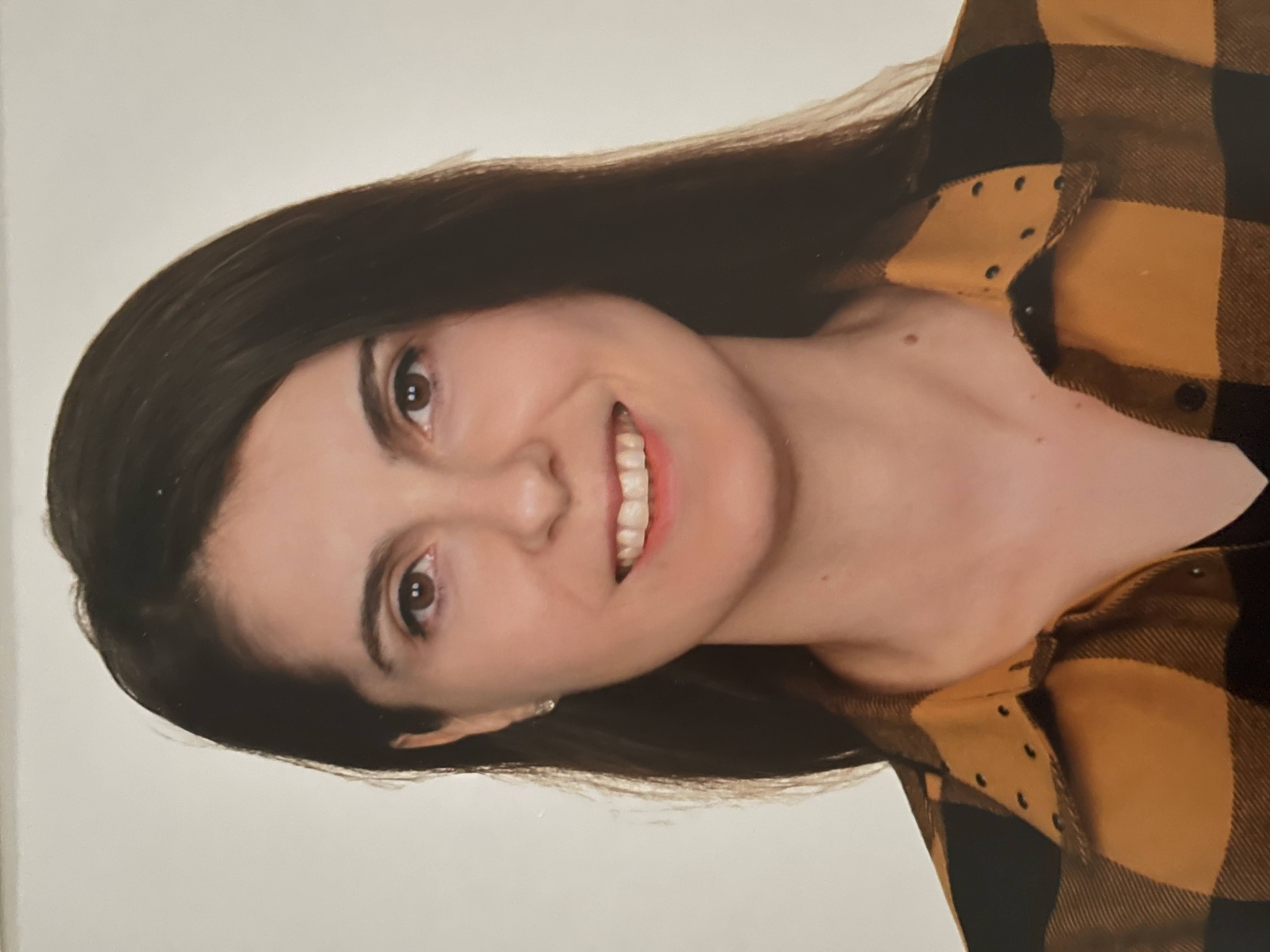}}
]
{Sevil Sen} is a full professor in the Department of Computer Engineering at Hacettepe University and leads the Wireless Networks and Intelligent Secure Systems (WISE) Laboratory. Her research interests include network and systems security, with a primary focus on mobile systems and wireless networks. She also serves as an Area Editor for Ad Hoc Networks and Genetic Programming and Evolvable Machines.
\end{IEEEbiography}

\end{document}